\documentclass[journal=jacsat,manuscript=article]{achemso}
\DeclareUnicodeCharacter{2082}{$_2$}
\usepackage[version=3]{mhchem} % Formula subscripts using \ce{}
\usepackage{hyperref}
\usepackage{xcolor, soul}
\author{Chhatra Bahadur Subba}
\affiliation[Mizoram University]
{Department of Physics, Mizoram University, Aizawl 796004, India}
\author{Bhanu Chettri}
\affiliation[Mizoram University]
{Department of Physics, Mizoram University, Aizawl 796004, India}
\author{Amel Laref}
\affiliation{Department of Physics and Astronomy, College of Science, King Saud University, Riyadh, 11451, Saudi Arabia}
\author{Zeesham Abbas}
\affiliation[GU1]{Department of Nanotechnology and Advanced Materials Engineering, Sejong University, Seoul, Republic of Korea}
\author{Amna Parveen}
\email{amnaparvin@gmail.com}
\affiliation[GU2]{College of Pharmacy, Gachon University, No. 191, Hambakmeoro, Yeonsu-gu, Incheon 21936, Republic of Korea}
\author{Dibya Prakash Rai}
\email{dibyaprakashrai@gmail.com}
\affiliation[Mizoram University]
{Department of Physics, Mizoram University, Aizawl 796004, India}
\author{Zaithanzauva Pachuau}
\affiliation[Mizoram University]
{Department of Physics, Mizoram University, Aizawl 796004, India}
\title[An \textsf{achemso} demo]
{Insights into the OER, ORR, and HER Activity of a New MXene-Family SnSiGeN$_4$ Photocatalyst for Water Splitting: A First-Principles Study}

\abbreviations{IR,NMR,UV}
\keywords{American Chemical Society, \LaTeX}

\begin{document}
	
	%\begin{tocentry}
	
	%\begin{center}
	%\includegraphics[scale=0.350]{structure.png}
	%\end{center}
	
	%\end{tocentry}
	
	\begin{abstract}
		The development of efficient and cost-effective catalysts for clean energy conversion remains a central challenge in materials science. Although platinum serves as the benchmark catalyst, its scarcity and high cost hinder large-scale deployment. In this study, we propose a newly predicted SnSiGeN$_4$ MXene-family monolayer as a promising candidate for the oxygen evolution reaction (OER), oxygen reduction reaction (ORR), and hydrogen evolution reaction (HER). Using first-principles calculations, we systematically investigated its electronic, vibrational, and optical properties across multiple exchange–correlation functionals, including hybrid approaches, revealing a wide and tunable band gap. Simulated infrared and Raman spectra further confirm the dynamical stability and the presence of catalytically active sites. Guided by these findings, we studied photocatalytic reaction analyses that demonstrate that the computed overpotentials for OER, ORR, and HER are comparable to those of Pt-based catalysts and outperform Ir-based systems, positioning SnSiGeN$_4$ as a sustainable, high-performance platform for next-generation UV–visible-light-driven photocatalysis.
	\end{abstract}
	
	\section{Introduction}
	The increasing global energy demand and the depletion of fossil fuels have intensified the search for clean and renewable energy sources. Among various sustainable energy conversion technologies, photocatalytic water splitting has attracted significant attention due to its potential for producing green hydrogen and oxygen without harmful emissions. The efficiency of photocatalysts largely depends on their electronic structure, light absorption ability, and charge carrier dynamics. In this context, two-dimensional (2D) materials have emerged as promising candidates owing to their large surface area, tunable band structures, and strong quantum confinement effects, which enhance catalytic activity\cite{tong2025interface,upadhyay2025enhanced,luo2016recent,almayyali2025new,su2020heterojunction}.
	
	The discovery of graphene\cite{geim2007rise} has inspired tremendous interest in two-dimensional (2D) materials due to their exceptional properties such as high charge mobility, tunable band gaps, and flexibility\cite{singh2019flexible,subba20252d,wang2015pseudocapacitance,subbacomprehensive,anayee2025layer,lemme20222d,gogotsi2023future}. These materials are now employed in devices including sensors, transistors, light-emitting diodes, quantum computing components, and photocatalysts\cite{wang2021enhanced,ullah2022mxene,schue2019bright,cobarrubia2024hexagonal,kuang2020mxene}. Continued exploration has led to the synthesis or prediction of silicene\cite{kara2012review}, phosphorene\cite{carvalho2016phosphorene}, germanene\cite{acun2015germanene}, and transition metal dichalcogenides (TMDs)\cite{manzeli20172d}. TMDs are particularly attractive owing to their tunable electronic and magnetic properties, ranging from semiconducting to metallic and even superconducting phases\cite{mak2010atomically,soluyanov2015type,liu2013three,liu2020robust,neto2001charge,han2023reversible}. Moreover, 2D g-C$_3$N$_4$ shows improved photocatalytic hydrogen generation of via efficient charge separation imposed by Bi$_2$O$_2$Se nanosheets \cite{lin2024improving}.
	
	Since 2011\cite{naguib2011two}, MXenes—a family of transition metal carbides, nitrides, and carbonitride with a general formula M$_{n+1}$X$_n$T$_x$ (where M represents an early transition metal, X = C or N, and T$_x$ denotes surface terminations such as –O, –OH, or –F)—have gained remarkable interest for energy-related applications\cite{subba2025comprehensive,kumar2022methods,ajmal2025advancements}. Their unique combination of metallic conductivity, chemical stability, and surface functionalization tunability makes them ideal for photocatalytic processes\cite{subba20252d,ajmal2025advancements}. Recent studies have shown that surface terminations play a crucial role in modulating the band gap, Fermi level, and charge transfer characteristics of MXenes\cite{nag2025modulation,mokkath2025crucial,bu2025tuning,doi:10.1021/acs.jpclett.0c03710}. Functionalization can transform metallic MXenes into semiconducting forms suitable for visible-light-driven photocatalysis. Moreover, heterostructure formation and defect engineering have been explored to further enhance the light absorption and charge separation efficiency in MXene-based composites\cite{subba20252d,prakash2025mxene,mahapatra2025artificial}. 
	When compared with other photocatalytic materials---including oxide-based catalysts~\cite{abdullah2025exploring,gupta2025review,yang2024photocatalytic}, metal chalcogenides~\cite{rao2025insights,guo2025customizing,mansoor2025comparative}, transition metal dichalcogenides (TMDs)~\cite{dange2025two,wijaya20252d,zhang2025design}, transition metal oxides (TMOs)~\cite{li2025modulation,zaera2025role,singh2025insight}, clay-based systems~\cite{ugwuja2025clay,ruiz2023mxenes,fatimah2022clay}, graphene~\cite{sadiq2025graphene,chu2025graphene,shafi2025facile,fauzia2024antibacterial}, layered double hydroxides (LDHs)~\cite{hu2025recent}, and metal oxide/sulfide photocatalysts such as MoS, MoS$_2$~\cite{katayama2025photocatalytic,lv2025molybdenum,das2024three}, WS$_2$~\cite{patil2024nanoscale,das2024improving}, CeO$_2$~\cite{herzog2024nanostructured,li2025cerium,fauzia2024antibacterial}, Bi$_2$S$_3$~\cite{chen2024improving,an2025photocatalytic}, CdS~\cite{shrestha2025lattice,bayram2025comparison,wu2025photocatalytic}, ZnS~\cite{bayram2025comparison}, V$_2$O$_5$~\cite{yuan2025novel,bravo2025enhancing}, Si$_3$N$_4$~\cite{nemaga2025unveiling}, ZnO~\cite{bayram2025comparison,xu2025improved}, TiO$_2$~\cite{deng2025synergistic,hiramatsu2024surface}, Nb$_2$O$_5$~\cite{mahmud2025visible,onur2024exploring}, Ta$_2$O$_5$~\cite{ngo2025photocatalytic}, WO$_3$~\cite{zhang2025tungsten,zhang2025carbon}, and SrTiO$_3$~\cite{davis2025unveiling}---MXenes display markedly superior photocatalytic activity~\cite{ranjith2024interfacial,yang2022rationally,zhong2021two}. This enhancement stems from their large surface-to-volume ratio, high electrical and optical conductivity, and tunable surface terminations achieved via wet-chemical etching~\cite{gogotsi2023future,murali2022review}.
	Despite these promising developments, a comprehensive understanding of the intrinsic electronic features that govern the photocatalytic behaviour of MXenes remains limited and pristine MXenes are good conductors with zero band gap. Many studies have primarily focused on experimental synthesis or composite fabrication, while detailed theoretical insights into band-gap engineering, band edge alignment, charge density distribution, and optical absorption mechanisms are still insufficient. Furthermore, the role of surface terminations and atomic composition on the electronic and photocatalytic properties needs systematic investigation at the atomic scale.
	
	Parallel to these developments, MXene-like 2D nitrides such as MoSiN$_4$ and WSiN$_4$ have been synthesized, where nitrogen substitutes for a chalcogen site in conventional TMDs\cite{hong2020chemical,chen2021first,mortazavi2021exceptional}. This discovery has expanded interest in the MA$_2$N$_4$ family (M = W, V, Nb, Ta, Cr, Mo, Zr, Hf; A = Si, Ge), which shows potential in energy storage, sensing, and catalysis\cite{dong2018metallic,lukatskaya2023ultra}. Among them, MoSi$_2$N$_4$ has been studied theoretically using PBE and HSE06 functionals\cite{heyd2005energy,heyd2003hybrid}, but remains experimentally unrealized.  
	In this study, we employ density functional theory (DFT) to recently report a new MXene family \cite{dat2022first} with a promising band gap in its pristine form to systematically investigate structural, electronic, vibrational, and optical properties of SnSiGeN$_4$ monolayer. The optical and vibrational spectra provide insights into its optical response and interaction with adsorbed species. Then we studied the SnSiGeN$_4$ monolayer as a bifunctional photocatalyst for OER/HER and OER/ORR reactions.
	
	\section{Computational Details}
	%\subsection{Periodic DFT Calculations}
	We performed all calculations at the level of unrestricted hybrid density functional theory, with long-range dispersion interactions treated using Grimme’s semi-empirical DFT-D3 correction. The equilibrium geometries, lattice parameters, and electronic properties—including band structure, band gap ($E_g$), and density of states (DOS)—of 2D monolayer SnSiGeN$_4$ were investigated using the CRYSTAL17 \textit{ab initio} code \cite{dovesi2017crystal17}. Periodic hybrid dispersion-corrected density functional theory (DFT) was employed for all computations, as implemented in the \textsc{CRYSTAL17} code, which utilizes localized Gaussian-type atomic orbitals (GTOs) as basis functions\cite{dovesi2014crystal14,bursch2018understanding}.
	In this calculations we have employed various exchange-correlation functionals to treat the electron interactions \cite{becke1988density,lee1988development,vosko1980accurate,perdew1996generalized,perdew2008restoring,vosko1980accurate,becke1993density,adamo1999toward,krukau2006influence,henderson2007importance,henderson2008assessment,weintraub2009long,zhao2008m06} including the most expensive (computationally) dispersion-corrected hybrid B3LYP-D3 functional \cite{grimme2010consistent,grimme2011effect,grimme2016dispersion}, which combines 20\% exact Hartree--Fock exchange with Grimme’s D3 correction for van der Waals interactions.  
	
	CRYSTAL17 utilizes localized Gaussian-type orbital (GTO) basis sets \cite{laun2022bsse}, which are more efficient than plane-wave methods for hybrid DFT. The pob\_TZVP\_rev2 triple-zeta polarized basis set was adopted \cite{laun2018consistent}. Brillouin-zone integrations were performed with Monkhorst--Pack meshes: $4 \times 4 \times 1$ for SnSiGeN$_4$ \cite{monkhorst1976special}. To study catalytic reactions, OER, ORR and HER, a $2 \times 2 \times 1$ SnSiGeN$_4$ supercell (4 Sn, 4 Ge, 4 Si and 16 N) and vacuum slabs were introduced to eliminate spurious interlayer interactions. 
	Convergence thresholds were set to $10^{-7}$ a.u. for energy, forces, and charge density and $10^{-10}$ a.u. for Gibbs Free energy calculation\cite{pascale2004calculation,zicovich2004calculation}. Band structures were computed along the high-symmetry $\Gamma$--M--K--$\Gamma$ path of the Brillouin zone, and DOS was obtained at the equilibrium geometries. Electrostatic potentials were referenced to the vacuum level. Structural models and charge densities were visualized using VESTA \cite{momma2008vesta}.  
	
	\subsection{Theoretical Framework for Reaction Free Energy Calculations}
	A $2 \times 2$ supercell of both the periodic 2D pristine SnSiGeN$_4$ monolayer and the possible active sites was employed in this study to investigate all the elementary steps of the Oxygen evolution reaction(OER), oxygen reduction reaction (ORR) and Hydrogen Evolution reaction(HER).
	
	\subsubsection{OER, ORR and HER Mechanisms}
	
	The oxygen evolution reaction (OER) on the surface of the catalyst is as follows:
	\begin{align}
		\text{H}_2\text{O}(l) + \,^* &\rightarrow \text{OH}^* + \text{H}^+ + e^- \\
		\text{OH}^* &\rightarrow \text{O}^* + \text{H}^+ + e^- \\
		\text{O}^* + \text{H}_2\text{O}(l) &\rightarrow \text{OOH}^* + \text{H}^+ + e^- \\
		\text{OOH}^* &\rightarrow \,^* + \text{O}_2(g) + \text{H}^+ + e^-,
	\end{align}
	where $*$ denotes the active site.  
	
	The four-electron transfer mechanism of ORR can be given as follows:
	\begin{align}
		\,^* + \text{O}_2(g) &\rightarrow 2\text{O}^* \\
		2\text{O}^* + \text{H}^+ + e^- &\rightarrow \text{OH}^* + \text{O}^* \\
		\text{OH}^* + \text{O}^* + \text{H}^+ + e^- &\rightarrow \text{O}^* + \text{H}_2\text{O} \\
		\text{O}^* + \text{H}^+ + e^- &\rightarrow \text{OH}^* \\
		\text{OH}^* + \text{H}^+ + e^- &\rightarrow \text{H}_2\text{O} + \,^*
	\end{align}
	
	The interactions between the intermediates and the 2D
	monolayer SnSiGeN$_4$ were estimated through the adsorption energy,
	as given by Eq.~\ref{eqads}:
	\begin{equation}
		E_{\text{ads}} = E_{\text{total}} - \left(E_{\text{sub}} + E_{\text{int}}\right),
		\label{eqads}
	\end{equation}
	where $E_{\text{total}}$ is the total energy of the 2D monolayer SnSiGeN$_4$ + OER/ORR/HER intermediate, $E_{\text{sub}}$ is the energy of the 2D monolayer SnSiGeN$_4$, and $E_{\text{int}}$ is the energy of the OER/ORR/HER intermediates. The zero-point energy ($E_\text{ZPE}$) and entropy ($S$) contributions were obtained from harmonic vibrational analysis as implemented in the \textsc{CRYSTAL17} code\cite{pascale2004calculation,zicovich2004calculation}. The contribution from the specific heat capacity was included in the Gibbs free energy ($G$) calculations; however, the lattice heat capacity $C_p$ was considered constant at the reference temperature of 298.15~K.

	The Gibbs free energy of each species can be expressed using Eq.~\ref{eqfreeEnerg}:
	\begin{equation}
		G = E_{\text{DFT}} + E_{\text{ZPE}} + \int C_{p}\, dT - TS
		\label{eqfreeEnerg}
	\end{equation}
	where $E_{\text{DFT}}$, $E_{\text{ZPE}}$, and $S$ are the ground state energy, 
	zero-point vibrational energy, and entropy, respectively. $T$ is taken to be 298.15~K.
	
	At constant temperature, the equation reduces to
	\begin{equation}
		G = E_{\text{DFT}} + E_{\text{ZPE}} - T S,
		\label{freeE}
	\end{equation}
	
	The change in free energy of reaction intermediates (adsorbates) was
	calculated at pH = 0 using
	\begin{equation}
		\Delta G_{\text{ads}} = E_{\text{ads}} + \Delta E_{\text{ZPE}} - T \Delta S,
		\label{Gads}
	\end{equation}
	where $\Delta G_{\text{ads}}$ is the adsorption free energy of adsorbates,
	and $\Delta E_{\text{ZPE}}$ and $\Delta S$ are the differences in
	zero-point energy and entropy, respectively.  
	
	The overpotential that evaluates the performance of the OER and ORR is defined as
	\begin{equation}
		\eta_{\text{OER}} = \frac{\max\{\Delta G_i\}}{e} - 1.23,
		\label{OERover}
	\end{equation}
	
	\begin{equation}
		\eta_{\text{ORR}} = \frac{\max\{\Delta G_i\}}{e} + 1.23,
		\label{ORRover}
	\end{equation}

	where $\Delta G_i$ represents the free energy change of each
	reaction step, and $e$ is the elementary charge.  
	
	The overall HER reactions using the photogenerated photoelectron can be written as
	% Photoexcitation
	\begin{equation}
		\text{Catalyst} + h\nu \;\rightarrow\; e^-_{\mathrm{CB}} + h^+_{\mathrm{VB}}
		\label{eqPhoAb}
	\end{equation}
	\begin{equation}
		^* \,+ \mathrm{H}^+ + e^-_{\mathrm{CB}} \;\rightarrow\; \mathrm{H}^*
		\label{actHyd}
	\end{equation}
	\begin{equation}
		\mathrm{H}^* \;\rightarrow\; \mathrm{1/2\,H}_2(g) + \,^*
		\label{H2Evol}
	\end{equation}
	\noindent
	where $*$ denotes the active site.
	
	\section{Results and discussion}
	\subsection{Structural Properties}
	\begin{figure}[H]
		\centering
		\includegraphics[width=\textwidth]{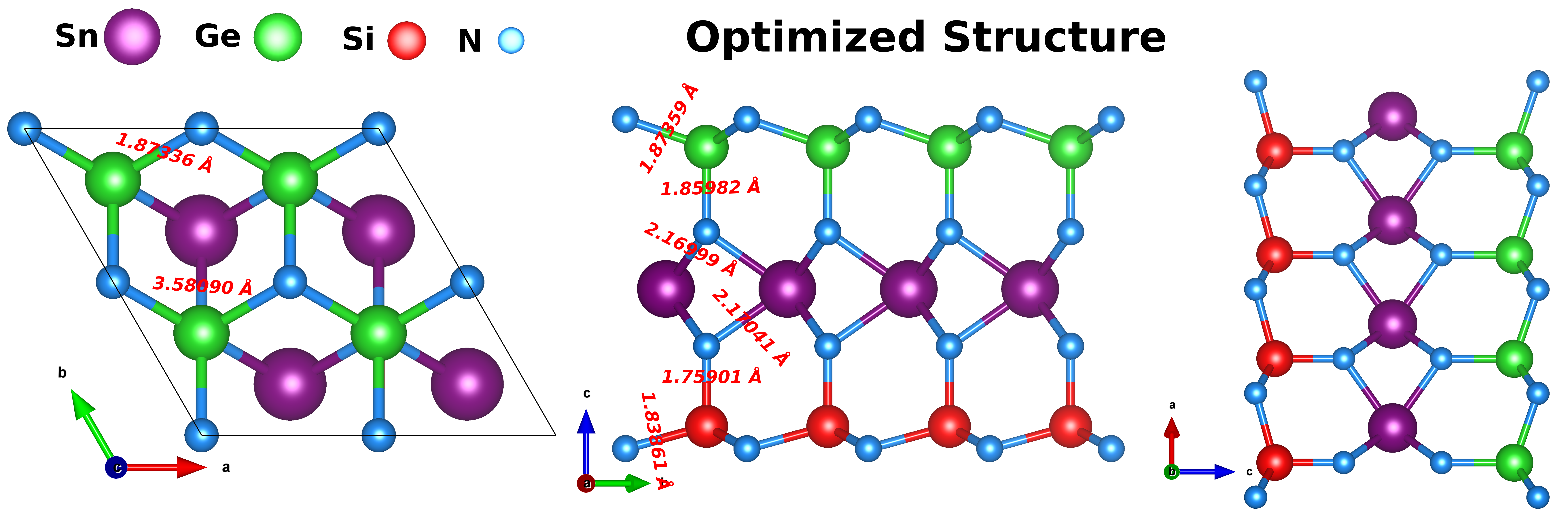}
		\caption{Top and side views of the optimized SnSiGeN$_4$ monolayer. The structure belongs to the hexagonal lattice system with space group P1 (No. 1). Atoms are labelled, and red-colored lines represent the calculated bond lengths between adjacent atoms.}
		\label{structure}
	\end{figure}
	To support its potential experimental realization and for real-world application, 
	the initial atomic structure of SnSiGeN$_4$ was modeled using the space group \textit{P1}, 
	in contrast to the earlier study which used \textit{P3m1}\cite{dat2022first}. Full structural optimization was 
	carried out using various DFT functionals within the CRYSTAL17 framework. The optimized 
	in-plane lattice constants were found to be in the range of $a = b = 3.0381$--$3.0915$~\AA, 
	depending on the chosen functional. A large out-of-plane lattice parameter 
	($c = 20$~\AA) was used to eliminate interlayer interactions between periodic images.
	
	Compared to earlier work\cite{dat2022first}, we observed a slight overestimation of the direct band gap 
	in HSE06 by $0.53$~eV and a difference from PBE by $0.05$~eV, which could be attributed to the 
	use of a different space group and a lower convergence tolerance ($10^{-4}$) in the previous 
	study. The detailed results of the geometric optimizations are summarized in 
	Table~\ref{table1}.
	
	\begin{table}[h]
		\caption{Structural and electronic properties of SnSiGeN$_4$ monolayer calculated 
			using different DFT functionals within the CRYSTAL17 code. Reported values include 
			the total energy, band gap, lattice parameters ($a$, $b$, $c$), and lattice angles 
			($\alpha$, $\beta$, $\gamma$).}
		\label{table1}
		\centering
		\resizebox{0.9\textwidth}{!}{   % <-- shrinks table to 90% of textwidth
			\begin{tabular}{lcccccccc}
				\hline
				Functional   & Energy (Hatree) & Direct Band Gap (eV) & $a$ (\AA) & $b$ (\AA) & $c$ (\AA) & $\alpha$ ($^\circ$) & $\beta$ ($^\circ$) & $\gamma$ ($^\circ$) \\
				\hline
				GGA (BLYP)   & -76192.7378 & 1.27 & 3.1262 & 3.1265 & 19.5753 & 90.0009 & 89.9947 & 120.0039 \\
				GGA (PBESOL) & -76101.3720 & 1.64 & 3.0662 & 3.0666 & 19.6552 & 90.0012 & 89.9979 & 120.0030 \\
				GGA (PBE)    & -76101.3720 & 1.50 & 3.0915 & 3.0919 & 19.7016 & 90.0009 & 89.9962 & 120.0040 \\
				GH (B3LYP)   & -76187.1955 & 2.98 & 3.0888 & 3.0890 & 19.6108 & 90.0013 & 89.9970 & 120.0030 \\
				GH (M06)     & -76192.4003 & 3.65 & 3.0570 & 3.0570 & 19.5719 & 90.0012 & 89.9930 & 120.0000 \\
				GH (PBE0)    & -76178.0892 & 3.58 & 3.0574 & 3.0576 & 19.5772 & 90.0014 & 89.9973 & 120.0027 \\
				RSH (HSE06)  & -76177.9611 & 2.92 & 3.0594 & 3.0596 & 19.5855 & 90.0014 & 89.9968 & 120.0032 \\
				RSH (HISS)   & -76181.0914 & 3.91 & 3.0381 & 3.0384 & 19.2405 & 90.0023 & 89.9989 & 120.0018 \\
				mGGA (M06L)  & -76198.7153 & 1.93 & 3.0658 & 3.0663 & 19.6298 & 90.0009 & 90.0143 & 120.0056 \\
				LDA (SVWN)   & -76055.4133 & 1.79 & 3.0463 & 3.0466 & 19.5824 & 90.0012 & 89.9974 & 120.0037 \\
				\hline
			\end{tabular}
		}
	\end{table}

	\subsection{Electronic Properties}
	
	To comprehensively evaluate the electronic and structural characteristics of 
	SnSiGeN$_4$ monolayer, geometry optimizations and band gap calculations were performed 
	using various exchange--correlation functionals within the CRYSTAL17 DFT framework. 
	Table~\ref{table1} summarizes the total energies, band gaps, and key lattice parameters 
	obtained from each functional. All functionals consistently yield a hexagonal unit cell 
	with lattice angle $\gamma \approx 120^\circ$, and $\alpha, \beta \approx 90^\circ$, 
	indicating preservation of the 2D layered structure. The in-plane lattice parameter 
	$a$ shows modest variation, ranging from $3.0381$~\AA\ (HISS) to $3.0915$~\AA\ (PBE). 
	The total energies are closely clustered between $-7.6055$~eV and $-7.6198$~eV, with 
	meta-GGA (M06L) predicting the most stable configuration.
	
	The predicted electronic band gaps vary significantly depending on the functional, 
	highlighting the sensitivity of electronic properties to the choice of exchange--correlation 
	treatment. It is well established that standard GGA functionals, such as PBE, tend to 
	underestimate band gaps in semiconductors and insulators due to their inadequate treatment 
	of exchange--correlation effects. To improve the accuracy of electronic structure predictions, 
	hybrid functionals like HSE, B3LYP, and PBE0 incorporate a fraction of exact Hartree--Fock 
	exchange, resulting in more reliable band gap estimations~\cite{dat2022first,muscat2001prediction,heyd2005energy}. 
	The semi-local functionals (e.g., LDA, PBE, BLYP) estimate band gaps in the range of 
	$1.27$--$1.79$~eV, while hybrid functionals (PBE0, HSE06, B3LYP, HISS) predict larger 
	gaps, reaching up to $3.91$~eV. Notably, LDA yields a moderate gap of $1.79$~eV. PBE and 
	PBESOL underestimate the gap, giving $1.49$~eV and $1.63$~eV, respectively.
	
	The Minnesota 2006 functional (M06L) and Henderson--Izmaylov--Scuseria--Savin (HISS) 
	functional yield significantly larger gaps of $3.65$~eV and $3.91$~eV respectively. 
	
	Overall, hybrid and meta-GGA functionals provide a more accurate representation for 
	photocatalysis-related studies, as they account better for electron localization and 
	exchange interactions. However, these results still await experimental validation.
	
	\begin{figure}[h]
		\centering
		\includegraphics[width=\textwidth]{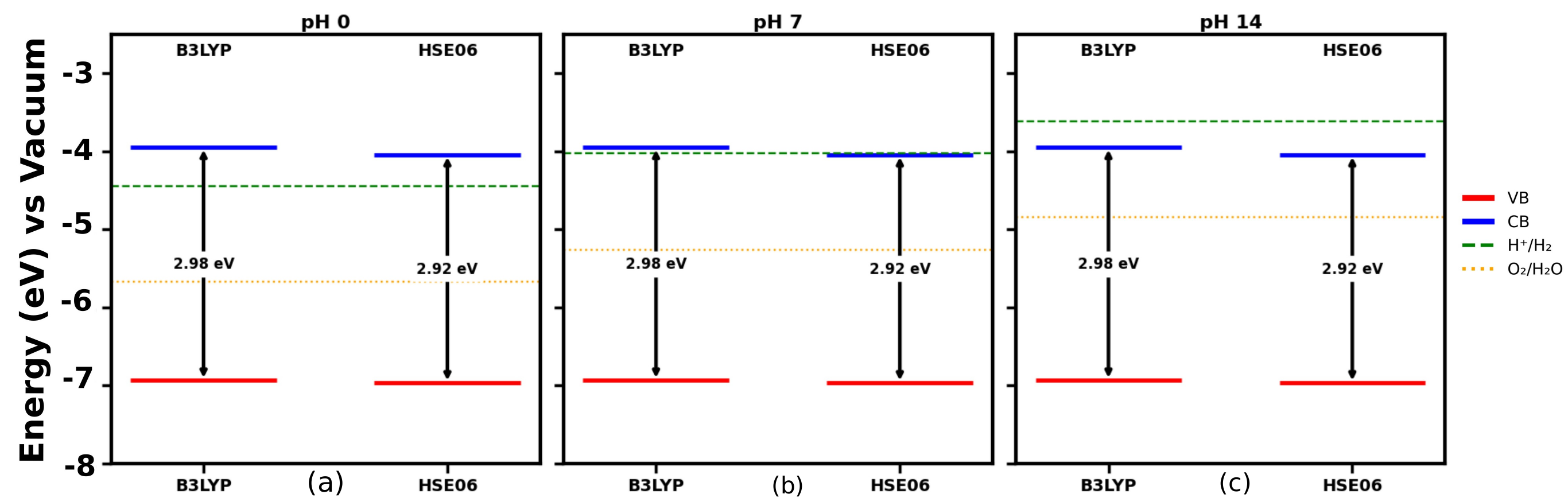}
		\caption{Schematic Representation of Band Edge Allignment at different pH.}
		\label{bandEdge}
	\end{figure}
	
	The band edge alignment in Fig.~\ref{bandEdge} demonstrates its strong potential as a photocatalyst for overall water splitting. The material exhibits a wide band gap of approximately 2.98~eV (B3LYP) and 2.92~eV (HSE06), which remains nearly invariant across pH values from 0 to 14. The conduction band minimum (CB) lies above the H$^{+}$/H$_{2}$ reduction potential (green dashed line), confirming the feasibility of proton reduction and efficient hydrogen evolution. Simultaneously, the valence band maximum (VB) is positioned below the O$_{2}$/H$_{2}$O oxidation potential (orange dotted line), enabling oxygen evolution under illumination. These favourable band edge positions, maintained over a wide pH range, indicate that SnSiGeN$_{4}$ fulfils the energetic requirements for spontaneous redox reactions, underscoring its suitability for photocatalytic water splitting applications and its promising electronic and catalytic properties.
	
	\begin{figure}[h]
		\centering
		\includegraphics[width=\textwidth]{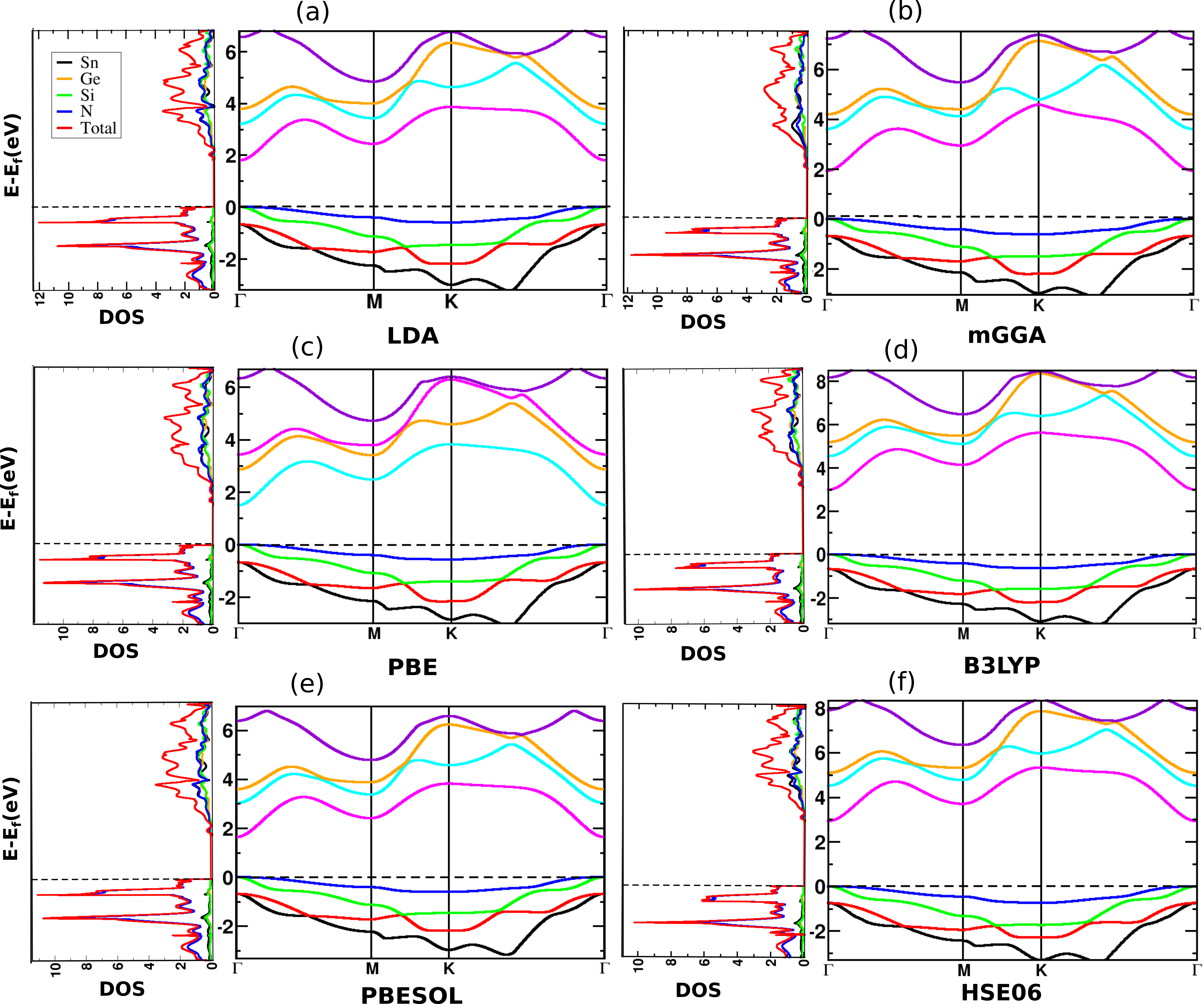}
		\caption{Projected density of states (PDOS) and electronic band structures of SnSiGeN$_4$ monolayer calculated using different exchange--correlation functionals.}
		\label{Electronic}
	\end{figure}
	
	Fig.~\ref{Electronic} shows the calculated electronic band structures and projected density of states (PDOS) for the SnSiGeN$_4$ monolayer obtained using various exchange--correlation functionals. Across all panels, the material consistently exhibits a direct band gap character. The calculated direct band gaps range from $1.27$~eV (BLYP) to $3.90$~eV (HISS), depending on the functional employed, consistent with the well-known underestimation of band gaps by semi-local functionals (e.g., LDA: $1.79$~eV, PBE: $1.50$~eV) and the improved predictions of hybrid functionals (e.g., HSE06: $2.92$~eV, PBE0: $3.58$~eV, HISS: $3.90$~eV).
	
	In the band structures, the valence band maximum (VBM) and conduction band minimum (CBM) appear at the $\Gamma$ points, leading to a direct band gap. This is consistent across all functionals studied. The valence bands are relatively flat near the VBM, suggesting a large hole effective mass, which may limit hole mobility. In contrast, the moderate dispersion of the conduction bands indicates comparatively lighter electrons, which could support better electron transport.
	In the previous studies\cite{dat2022first} show that the SnSiGeN$_4$ monolayer possesses very low and isotropic electron effective masses (0.21~$m_0$) with exceptionally high electron mobility ($\sim$675~cm$^2$~V$^{-1}$~s$^{-1}$), exceeding that of other MX$_2$N$_4$ (M = Mo, W; X = Ge, Si) compounds and making it excellent for charge separation. Its lower hole mobility ($\sim$51~cm$^2$~V$^{-1}$~s$^{-1}$), together with a high in-plane elastic modulus ($\sim$228~N/m) and moderate deformation potential constants, indicates robust mechanical stability and lattice tolerance, which are critical for efficient photocatalytic performance. Notably, Sn reverses the typical charge-carrier mobility trend observed in the MoSi$_2$N$_4$ family and related Sn-based monolayers, highlighting SnSiGeN$_4$ as a highly promising photocatalyst.\cite{mortazavi2021exceptional,tian2021electronic,dat2022layered,yao2021novel}
	
	The PDOS analysis reveals that the N $p$-states dominate both the VBM and CBM, with additional contributions from Ge and Si $p$-states near the valence band region. The Sn $s$-states are located deeper in the valence band, consistent with their role in structural stability rather than frontier electronic states. The total DOS consistently shows a clean separation of valence and conduction bands, confirming the semiconducting nature of SnSiGeN$_4$.
	
	A key observation is the widening of the band gap with hybrid functionals (e.g., HSE06, HISS) compared to LDA and GGA functionals. This trend reflects the improved treatment of exchange--correlation effects by including a fraction of Hartree--Fock exchange, which is crucial for accurate band gap prediction in 2D semiconductors. For instance, HSE06 yields a band gap of $2.92$~eV, significantly larger than PBE’s $1.50$~eV, aligning with expectations for better band gap estimation using hybrid methods.
	
	Overall, these electronic structure results confirm that SnSiGeN$_4$ is a robust direct semiconductor with a band gap tunable via the choice of functional, dominated by N-derived frontier orbitals. These properties make it a potential candidate for photocatalytic applications, where an appropriately sized band gap and stable electronic structure are essential.
	
	\begin{table}[h!]
		\centering
		\caption{Key Mulliken overlap populations (OVPOP) and bond lengths for O-, OH-, and OOH-adsorbed species for Sites V and VI. Only significant interactions (OVPOP $\geq$ 0.05) are included.}
		\begin{tabular}{l l r r}
			\hline
			\textbf{Adsorbate / Site} & \textbf{Bond} & \textbf{R (Å)} & \textbf{OVPOP} \\
			\hline
			OH* (Site V)  & O--H  & 0.986 & 0.208 \\
			& O--Si & 1.784 & 0.206 \\[1mm]
			OH* (Site VI) & O--H  & 0.984 & 0.237 \\
			& O--Si & 2.001 & 0.102 \\[1mm]
			O* (Site V)   & O--Si & 1.770 & 0.142 \\[1mm]
			O* (Site VI)  & O--Si & 1.761 & 0.129 \\[1mm]
			OOH* (Site V) & O--H  & 1.030 & 0.105 \\
			& O--Si  & 2.024 & 0.085 \\[1mm]
			OOH* (Site VI)& O--H  & 0.994 & 0.250 \\
			& O--Si & 1.939 & 0.118 \\
			\hline
		\end{tabular}
		\label{OVPOP}
	\end{table}

	From Table \ref{OVPOP}, the OER proceeds via stepwise formation of OH*, O*, and OOH* intermediates. 
	OH* adsorption is stabilized by strong O--H and O--Si bonds (OVPOP $\approx$ 0.208--0.237). 
	O* shows strong O--Si bonding (OVPOP $\approx$ 0.129--0.142) with moderate interaction with neighbouring N, allowing desorption. 
	OOH* formation is facilitated by strong O--H (OVPOP $\approx$ 0.105--0.250) and moderate O--O bonding (OVPOP $\approx$ 0.104), indicating feasible O--O coupling. 
	These features suggest favourable thermodynamics and kinetics for OER.
	
	The ORR, proceeding in reverse, involves O$_2$ adsorption and proton--electron transfer to form OOH*, OH*, and H$_2$O.  For OH* adsorption, Site V exhibits stronger O--Si bonding (1.784~\AA, OVPOP~0.206) compared to Site VI (2.001~\AA, OVPOP~0.102), indicating more stable initial anchoring. In the case of O*, both sites show comparable stabilization, with Site V (1.770~\AA, OVPOP~0.142) slightly stronger than Site VI (1.761~\AA, OVPOP~0.129). A significant difference emerges for OOH* adsorption, Site VI displays stronger O--H (0.994~\AA, OVPOP~0.250) and O--Si interactions (1.939~\AA, OVPOP~0.118) than Site V, where the O--Si interaction is notably weaker (2.024~\AA, OVPOP~0.085). These results suggest that while Site V favours the initial adsorption of OH*, Site VI provides more effective stabilization of the OOH* intermediate, which is often the potential-determining step in ORR. Thus, both Site V and VI can be considered the more active sites for facilitating ORR.
	
	OOH* is stabilized by strong O--H (OVPOP $\approx$ 0.25) and moderate O--Si (0.118) interactions, supporting low-overpotential pathways. 
	OH* and O* intermediates are sufficiently stabilized for efficient proton--electron transfer and product desorption.
	Overall, the bonding analysis shows that key intermediates for OER and ORR are stabilized optimally. Combined with strong visible-light absorption, these results indicate that the material is a promising bi-functional photocatalyst.

	\subsection{Optical and Vibrational Properties}
	\begin{figure}[h]
		\centering
		\includegraphics[width=\textwidth]{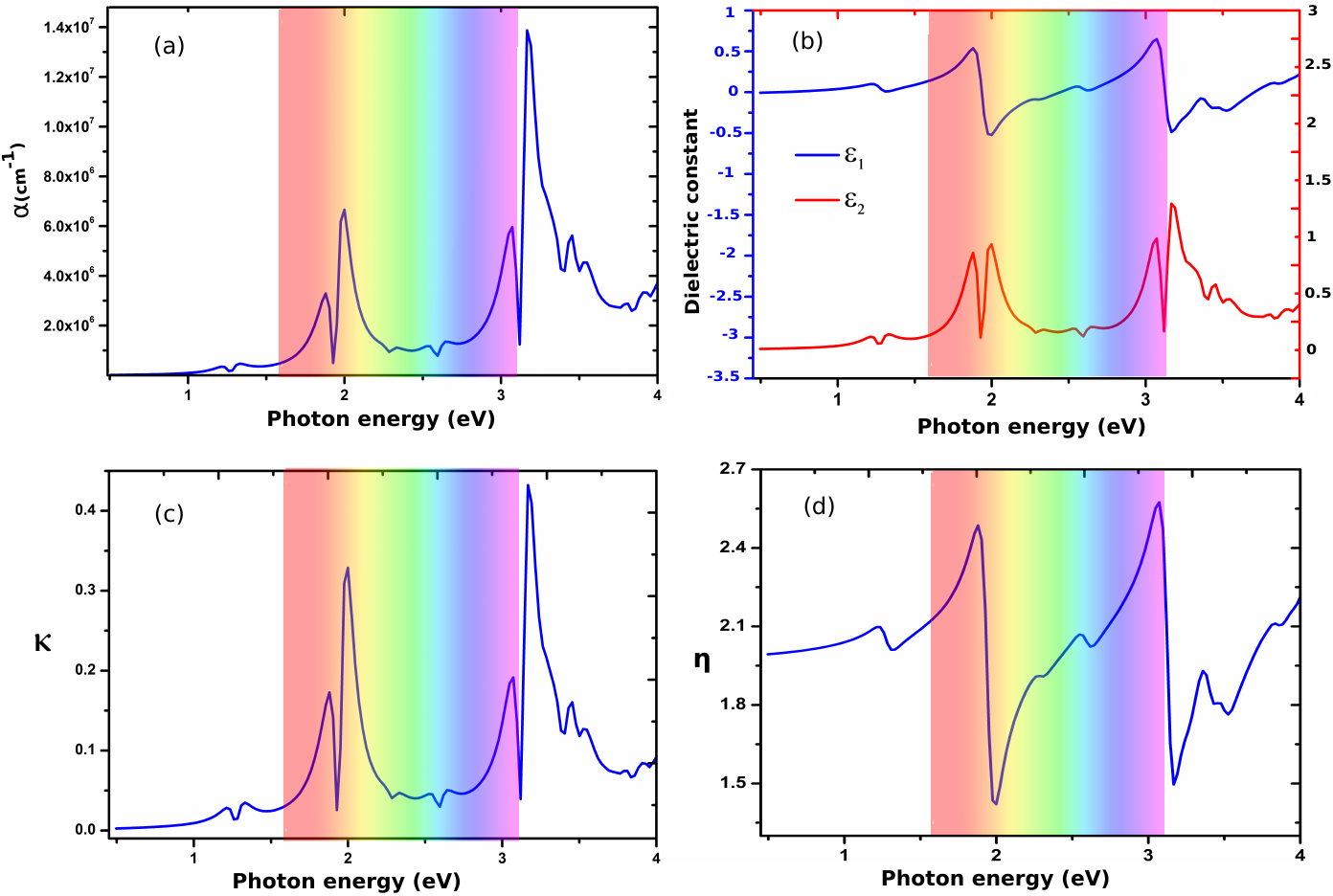}
		\caption{(a) absorption coefficient, (b) Dielectric constant, (c) extinction function, (d) refractive index.}
		\label{opticl}
	\end{figure}

	\begin{figure}[h]
		\centering
		\includegraphics[width=\textwidth]{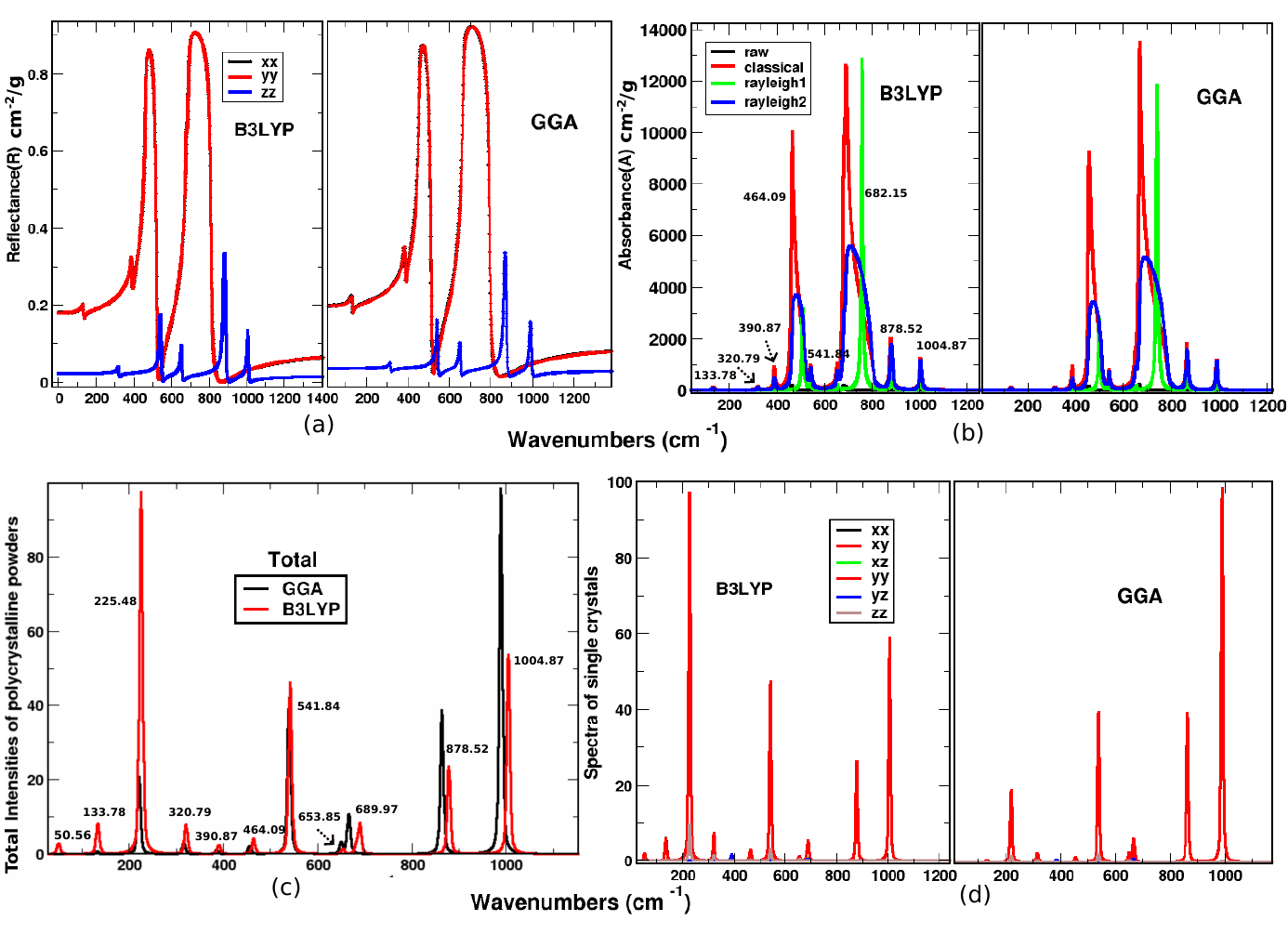}
		\caption{Vibrational IR and Raman spectra of the SnSiGeN$_4$ monolayer calculated with B3LYP and GGA functionals. The numerical label denotes the respective frequency mode.}
		\label{Raman}
	\end{figure}
	
	Fig~\ref{opticl}(a)-(d) shows the calculated optical properties of the material across the photon energy range, with the visible spectrum highlighted between 1.8~eV and 3.1~eV. Two sharp resonances near 2.0~eV and 3.0~eV are clearly seen in the absorption coefficient ($\alpha$) and extinction coefficient ($\kappa$) spectra, and are closely mirrored by peaks in the imaginary part of the dielectric function ($\varepsilon_2$), indicating efficient photon absorption.
	
	The real part of the dielectric function ($\varepsilon_1$) and the refractive index ($\eta$) exhibit anomalous dispersion near these resonances. Notably, $\varepsilon_1$ becomes negative beyond 2.5~eV, reaching $\sim -3.0$, suggesting surface plasmon resonances or intense interband transitions that enhance light-matter interaction.
	
	From a photocatalytic perspective, the strong visible-light absorption and well-defined dielectric response ($\varepsilon_1$ and $\eta \approx 1.8$) indicate efficient photon harvesting, favourable charge separation, and minimal energy loss. These combined features highlight the material’s potential as a visible-light-driven photocatalyst for water splitting.
	
	Next is to examine how vibrational properties affect the active sites, we examined both the IR and Raman spectra. These were calculated using the Coupled-Perturbed Hartree–Fock/Kohn–Sham (CPHF/KS) approach\cite{ferrero2008coupled,ferrero2008calculation}, which provides a direct method for computing Born charges and vibrational intensities. As shown in Figure~\ref{Raman}, the resulting vibrational spectra exhibit distinct features 
	that highlight the strong photocatalytic potential of SnSiGeN$_4$.
	
	The reflectance spectra [see Fig.~\ref{Raman}(a)] exhibit pronounced anisotropy between the in-plane (XX and YY) and out-of-plane (ZZ) polarization components, with strong peaks observed in the XX and YY directions at approximately $450~\text{cm}^{-1}$ and $750~\text{cm}^{-1}$, respectively, where the reflectance values exceed $0.8$. In contrast, the ZZ component remains negligible across the examined spectral range. This pronounced optical anisotropy clearly indicates the two-dimensional nature of SnSiGeN$_4$.
	
	The infrared absorbance spectra [see Fig.~\ref{Raman}(b)] exhibit intense absorption features below $800~\text{cm}^{-1}$, with prominent peaks in the $450$--$650~\text{cm}^{-1}$ range, which can be attributed to lattice vibrational modes associated with Sn--N, Si--N, and Ge--N stretching vibrations. These strong IR-active phonon modes reflect the robust bonding and structural stability of SnSiGeN$_4$. Moreover, the inclusion of raw, classically corrected, and Rayleigh-corrected spectra validates the robustness of these absorption features against possible scattering effects.
	
	The Raman spectra [see Fig.~\ref{Raman}(c and d)] exhibit several sharp and well-resolved vibrational modes, with the most intense peaks below $300$~cm$^{-1}$, and weaker features extending up to $\sim 1000$~cm$^{-1}$. These modes are primarily attributed to Sn--N and Si/Ge--N bond vibrations, which are characteristic of layered nitride structures.  
	
	According to group theory, the triclinic \textit{P-1} space group (C$_1$ point group) of SnSiGeN$_4$ has no symmetry beyond identity, meaning all modes are of A symmetry and are both Raman and IR active. This explains the observation of Raman peaks in all polarization directions[see Fig.~\ref{Raman}(d)], consistent with the lack of symmetry constraints in low-symmetry systems. Comparing the B3LYP and GGA results, both approaches show qualitatively similar trends, although slight shifts ($5$--$20$~cm$^{-1}$) in peak positions and variations in relative intensities are evident. These differences arise from the exchange--correlation approximations inherent to each functional. The Raman spectra of polycrystalline [see Fig.~\ref{Raman}(c)] powders further demonstrate the experimental feasibility of detecting these characteristic modes, offering clear fingerprints for phase identification and material characterization.
	
	\begin{figure}[h]
		\centering
		\includegraphics[width=\textwidth]{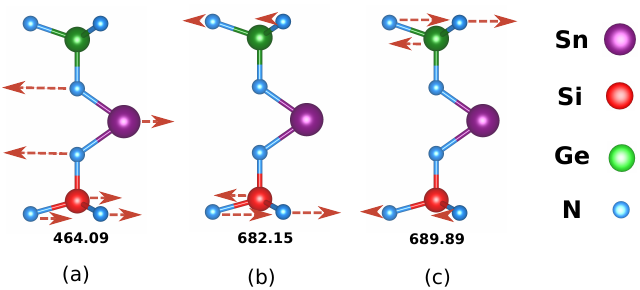}
		\caption{Representative IR active vibrational mode  in SnGeSiN$_4$which enhance the activation of catalytic sites 
			by facilitating intermediate adsorption and stabilization.}
		\label{mode}
	\end{figure}
	
	These distinct vibrational modes strongly correlate with catalytic functionalities. 
	The low-frequency mode at 464.09 cm$^{-1}$ (Sn--N/Ge--N stretching) is particularly favourable for the hydrogen evolution reaction (HER), as it provides dynamic hydrogen binding sites and facilitates H--H recombination. 
	In the mid-frequency region, the 653--700 cm$^{-1}$ range facilitates to * --O, * --OH, and * --OOH stretching vibrations(* denote active site), which play a crucial role in the oxygen reduction (ORR) and oxygen evolution reactions (OER) by enabling O$_2$ activation and stabilizing key intermediates such as OH* and OOH*. 
	Furthermore, the high-frequency modes above 878 cm$^{-1}$ enhance ORR and OER by promoting O--O bond cleavage during reduction and O--O bond formation during evolution. 
	Collectively, the coexistence of IR- and Raman-active modes across low, mid, and high frequency ranges highlights the multifunctional nature of SnGeSiN$_4$, establishing feasibility for HER, ORR, and OER.
	
	\subsection{Catalytic Properties}
\begin{figure}[h]
		\centering
		\includegraphics[width=\textwidth]{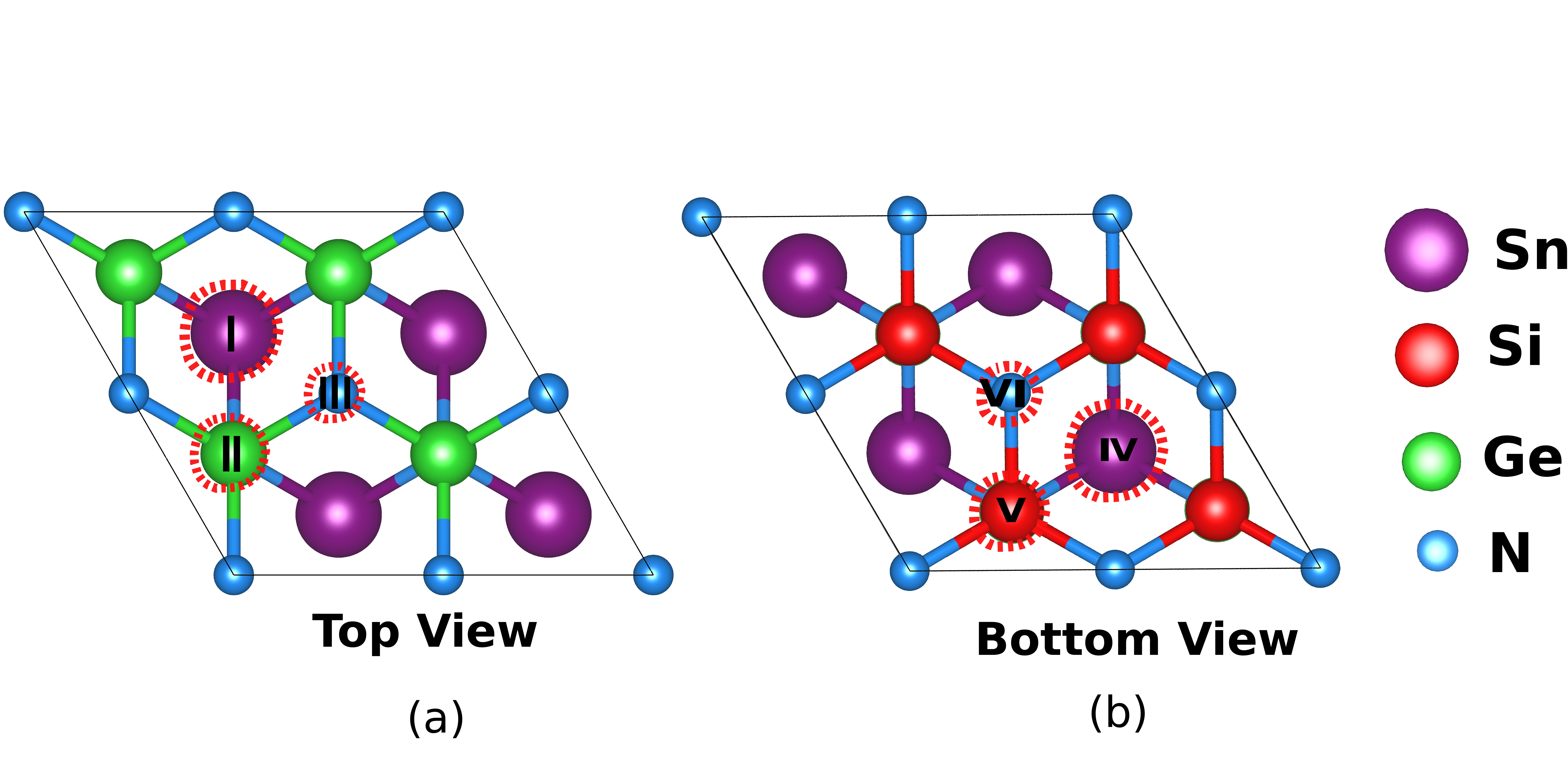}
		\caption{Different possible active sites for OER, ORR and HER reactions.}
		\label{activesite}
	\end{figure}

	With this compelling evidence (electronic, optical and vibrational analysis), we further investigated the oxygen evolution reaction (OER), oxygen reduction reaction (ORR), and hydrogen evolution reaction (HER), to demonstrate the catalytic activity of the subject material. Firstly, we choose six possible active sites as shown in Fig.~\ref{activesite}.
	The catalytic adsorption sites were defined based on the atomic environment of the layered structure. Sites I, II, and III are located on top of Sn, Ge, and N atoms of the upper layer, respectively, while Sites IV, V, and VI correspond to positions above Sn, Si, and N atoms of the bottom layer. The distinction between the two sets of sites arises from the different atomic constituents in the top and bottom layers, which create inequivalent chemical environments for catalytic activity. The adsorption geometries of reaction intermediates were further optimized to identify their most favourable binding sites, which are critical for evaluating the catalytic performance toward OER, ORR, and HER. For all these processes, the Gibbs free energy ($\Delta G$) was calculated using equations (\ref{eqads}-\ref{Gads}), representing the change in Gibbs free energy associated with each elementary reaction step.
	
	\subsubsection{OER Catalytic Activity}
	
	The oxygen evolution reaction (OER) was analyzed by calculating the Gibbs free energy change ($\Delta G$) associated with each elementary step of the reaction pathway. The process proceeds through four sequential proton--electron transfer steps involving adsorbed intermediates. To capture the energetics, the adsorption geometries of these intermediates were optimized, and the corresponding $\Delta G$ values were obtained from DFT-D calculations (Table~\ref{OER_free_energy}).
	
	\begin{table}[h!]
		\centering
		\caption{Gibbs Free Energy Change ($\Delta G$) for Various Reaction Intermediates during the OER Process on SnGeSiN$_4$ at Different Adsorption Sites (I–VI).}
		\label{OER_free_energy}
		\footnotesize
		\begin{tabular}{lcccccc}
			\hline
			\textbf{OER Step} & \textbf{Site I} & \textbf{Site II} & \textbf{Site III} & \textbf{Site IV} & \textbf{Site V} & \textbf{Site VI} \\
			\hline
			H$_2$O $\rightarrow$ OH$^*$ + (H$^+$ + e$^-$) & 2.3955 & 2.4213 & 2.4247 & 2.5858 & 1.7138 & 2.2217 \\
			OH$^*$ $\rightarrow$ O$^*$ + (H$^+$ + e$^-$) & 1.1414 & 1.2450 & 1.3558 & 0.2572 & 1.2024 & 0.4460 \\
			O$^*$ + H$_2$O $\rightarrow$ O$^*$ + OH$^*$ + (H$^+$ + e$^-$) & 0.6637 & 0.4393 & 0.3267 & 1.6841 & 1.0067 & 1.8680 \\
			O$^*$ + OH$^*$ $\rightarrow$ O$_2$ + (H$^+$ + e$^-$) & 0.7194 & 0.8145 & 0.8127 & 0.3928 & 0.9971 & 0.3842 \\
			\hline
		\end{tabular}
	\end{table}

	On the catalyst surface, the OER mechanism begins with the adsorption and dissociation of H$_2$O to form a hydroxyl species (OH*). This is followed by the deprotonation of OH* to generate adsorbed O*. The next step involves the interaction of O* with a water molecule to produce the OOH* intermediate, which subsequently evolves into molecular O$_2$ and desorbs from the surface. Each of these steps is associated with distinct free-energy changes, and their magnitudes dictate the overall reaction kinetics.
	\begin{figure}[H]
		\centering
		\includegraphics[width=\textwidth]{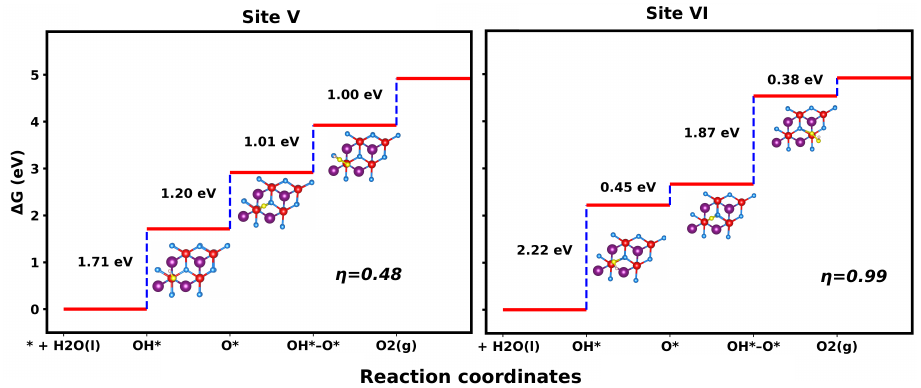}
		\caption{Gibbs free energy diagram of the OER pathway on the
			surface of the 2D monolayer SnSiGeN$_4$ with optimized structures.}
		\label{OER}
	\end{figure}
	Table~\ref{OER_free_energy} reveals that the most energetically demanding step ($\Delta G $) is the first step, i.e adsorption of H$_2$O on the catalyst, forming a hydroxyl species (OH*), thus serving as the rate-determining step for different sites for all six sites, which is the indicator for OER performance of the catalyst. The overpotentials calculated using equation \ref{OERover} for OER for site I to site VI are 1.16 eV, 1.19 eV, 1.19 eV, 1.35 eV, 0.48 eV and 0.99 eV, respectively.  Since all steps are uphill in energy, an external potential source from sunlight is required to drive the reaction. The calculated overpotential ($\eta$) at Site~V is 0.48~V, which is the lowest among six sites and is lower than that of the benchmark IrO$_2$ catalyst ($\eta = 0.56$~V) \cite{man2011universality} and close to PtO$_2$-rutile ($\eta = 0.4$~V) \cite{norskov2004origin}. Fig. \ref{OER} shows the Gibbs free energy diagram of OER pathways for sites V and VI, which have the lowest overpotential among the six, which is also predicted from vibrational analysis [ see Fig. \ref{mode}(b \& c)], are the most active sites. Considering the high cost and scarcity of iridium and platinum-based catalysts, the 2D SnSiGeN$_4$ system represents a cost-effective and efficient alternative with competitive catalytic performance.

	\subsubsection{ORR Catalytic Activity}
	\begin{figure}[H]
		\centering
		\includegraphics[width=\textwidth]{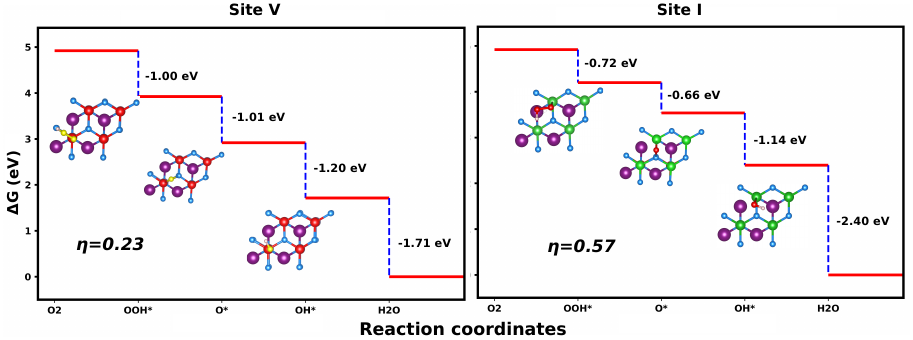}
		\caption{Gibbs free energy diagram of the ORR pathway on the
			surface of the 2D monolayer SnSiGeN$_4$ with optimized structures}
		\label{ORR}
	\end{figure}
	
	For the oxygen reduction reaction (ORR)—the reverse of the oxygen evolution reaction (OER)—we evaluated the relative Gibbs free energy ($\Delta G$) associated with each elementary reaction step to understand the reaction mechanism and catalytic performance. Under acidic conditions, ORR can proceed via two possible mechanisms: the associative pathway (O$_2$~$\rightarrow$~OOH) or the dissociative pathway (O$_2$~$\rightarrow$~2O). The distinction between these pathways depends on whether the adsorbed O$_2$ molecule dissociates into two separate O atoms upon interaction with the catalyst surface. If dissociation occurs, the reaction follows the dissociative mechanism; otherwise, it proceeds via the associative route, wherein one proton (H$^+$) and one electron (e$^-$) simultaneously adsorb onto the O$_2$ molecule, forming an OOH intermediate on the catalytic surface. Hence, Gibbs free energy analysis provides key insights into the thermodynamic preference of the reaction pathway on SnGeSiN$_4$.
	
	The free energy profile of the ORR at Site~V is illustrated in Fig.~\ref{ORR}. All the elementary steps are energetically favourable, indicating spontaneous feasibility under illumination. The calculated Gibbs free energy changes ($\Delta G$) for the successive reduction steps---O$_2$~$\rightarrow$~OOH*, OOH*~$\rightarrow$~O*, O*~$\rightarrow$~OH*, and OH*~$\rightarrow$~H$_2$O---are $-1.00$, $-1.01$, $-1.20$, and $-1.71$~eV, respectively. The corresponding overpotential ($\eta$) is determined to be $0.23$~eV, suggesting excellent catalytic activity toward the ORR outperforming Pt ($\eta = 0.4$~eV) \cite{norskov2004origin}.
	
	As illustrated in the free energy diagram (Fig.~\ref{ORR}), all electrochemical steps are exergonic (downhill), confirming the thermodynamic feasibility of the reaction. Among them, the protonation of O$_2$* to OOH* is identified as the potential-limiting step, with $\Delta G = -1.00$~eV. Thus, all reaction steps remain exothermic up to the limiting potential. According to Nørskov \textit{et al.}\cite{viswanathan2012unifying}, an ideal ORR catalyst has a $\Delta G$ of $-0.86$~eV for the OH*~$\rightarrow$~H$_2$O step. The calculated value for SnGeSiN$_4$ is better than this ideal, implying that the material lies near the apex of the 4$e^-$ ORR activity volcano.
	
	Furthermore, the optimized O--O bond length of the adsorbed O$_2$ species is found to be 1.23~\AA, with an adsorption energy ($\Delta E$) of $-0.41$~eV---closely matching that of the gas-phase O$_2$ molecule. This observation confirms molecular (non-dissociative) adsorption of O$_2$ on the surface. Together with the smooth downhill energy profile, these results indicate that the ORR on Site~V proceeds via the associative mechanism. The low overpotential and stable molecular adsorption make Site~V an energetically favourable active site for efficient oxygen reduction. The next most favourable site, Site~VI, also follows the associative mechanism but exhibits a slightly higher overpotential ($\eta = 0.57$~eV).
	
	\subsubsection{HER Catalytic Activity}
	The entire process of the hydrogen evolution reaction (HER) can be represented 
	by a three-state diagram, comprising the following states: 
	
	\begin{enumerate}
		\item Initial state: H$^+$ + e$^-$ (proton and electron),
		\item Intermediate state: H$^*$ (hydrogen atom in an activated state),
		\item Final product: H$_2$ (molecular hydrogen).
	\end{enumerate}

	The Gibbs free energy of hydrogen atom adsorption ($\Delta G_{\mathrm{H}^*}$) is a key 
	parameter for evaluating the catalytic activity. An ideal HER catalyst should exhibit 
	a $\Delta G_{\mathrm{H}^*}$ close to zero \cite{jiao2015design}, with the optimal value approaching zero. If $\Delta G_{\mathrm{H}^*}$ is too positive, hydrogen 
	adsorption on the surface is unfavourable. Conversely, if $\Delta G_{\mathrm{H}^*}$ is too 
	negative, hydrogen atoms bind too strongly, making their release as H$_2$ difficult. 
	
	Therefore, achieving a $\Delta G_{\mathrm{H}^*}$ near zero ensures a balanced adsorption and 
	desorption process, which is crucial for efficient hydrogen production. The closer the Gibbs free energy of hydrogen adsorption, $|\Delta G_{\mathrm{H}}|$, is to zero, the higher the catalytic activity for the hydrogen evolution reaction (HER).
	
	\begin{figure}[H]
		\centering
		\includegraphics[width=0.7\textwidth]{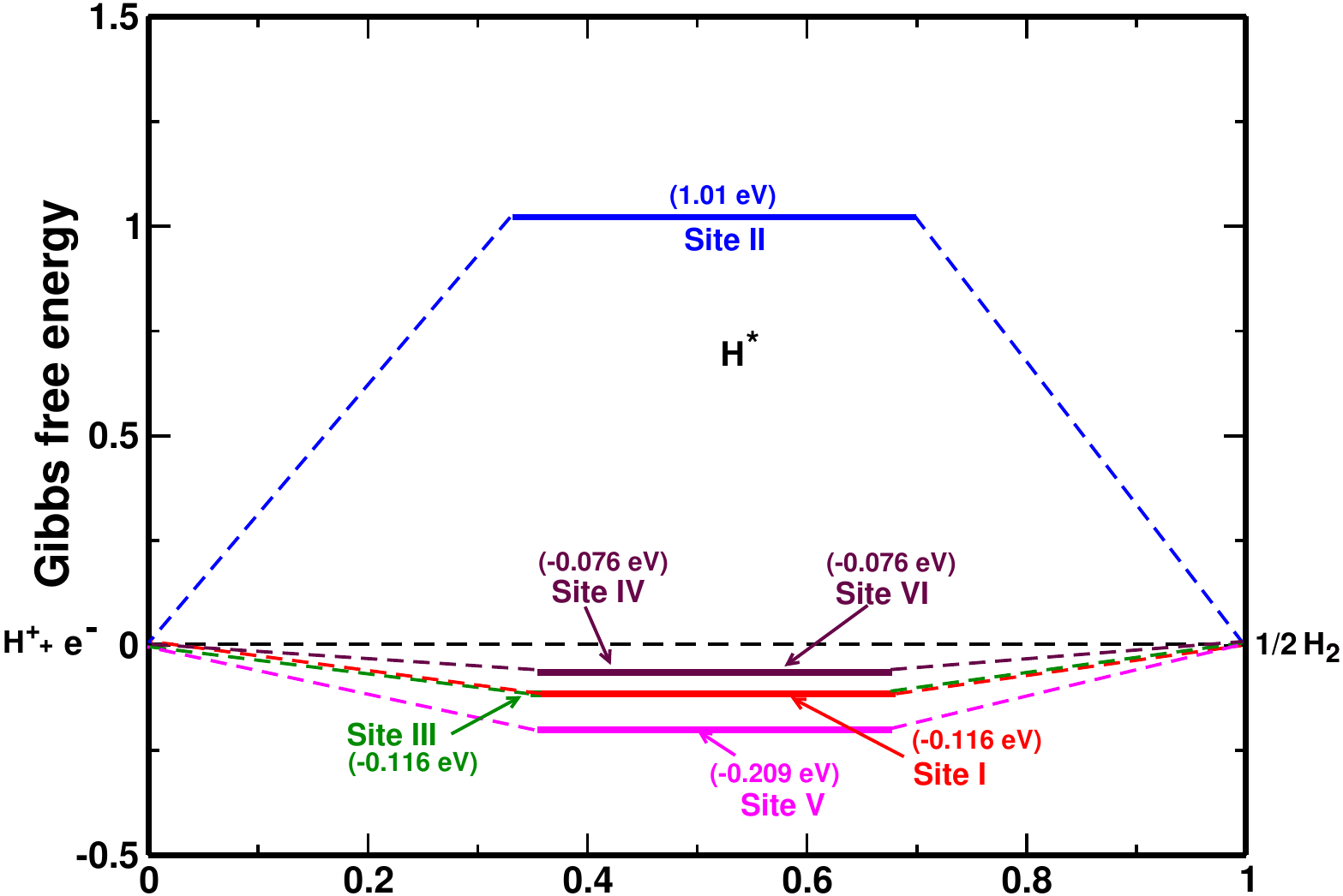}
		\caption{Calculated Gibbs free energy diagram for the HER at different sites of the pristine 2D monolayer}
		\label{Gibbs2}
	\end{figure}
	In our DFT calculations, pristine SnGeSiN$_4$ exhibits varying $|\Delta G_{\mathrm{H}}|$ values across different adsorption sites. The positive value at Site II ($\Delta G_{\mathrm{H}^*} = 1.01$~eV) indicates relatively loose binding of H$^*$, suggesting inefficient adsorption. In contrast, Site I and Site III display $|\Delta G_{\mathrm{H}^*}|$ values comparable to benchmark catalysts such as Pt (0.09~eV)\cite{norskov2005trends,hinnemann2005biomimetic}, MoS$_2$ (0.08~eV)\cite{bonde2009hydrogen}, and WS$_2$ (0.22~eV)\cite{bonde2009hydrogen}, indicating favorable hydrogen adsorption and enhanced HER activity. Interestingly, Sites IV and VI, which merge into a single site after structural optimization, exhibit a Gibbs free energy change close to zero—surpassing the catalytic performance of Pt, MoS$_2$, and WS$_2$ (0.22~eV), which is also supported by vibrational analysis [see Fig.~\ref{mode}(a)].
	\begin{figure}[h!]
		\centering
		\includegraphics[width=0.85\textwidth]{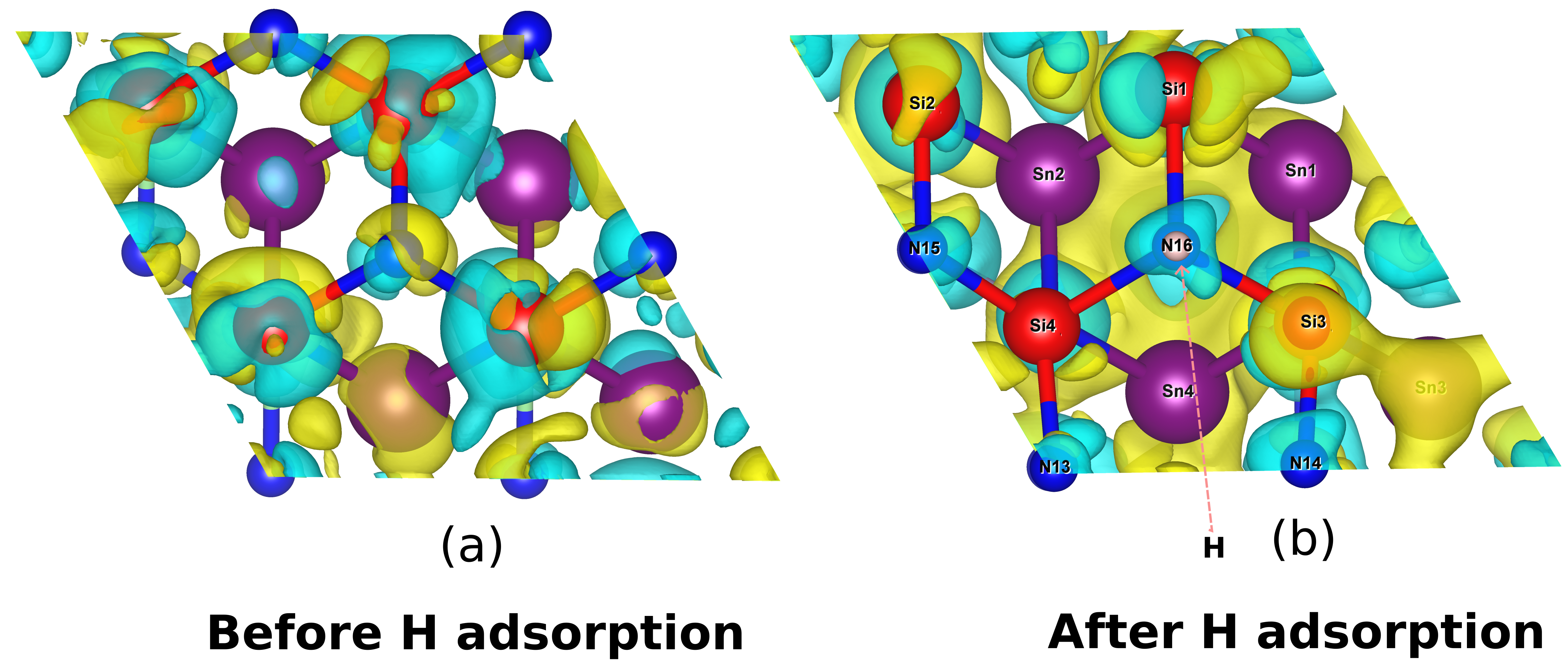}
		\caption{Spin density isosurfaces of SnGeSiN$_4$: (a) Before and (b) After H-adsorbed surface. 
			Yellow and cyan regions correspond to spin-up and spin-down electron density, respectively. 
			Upon hydrogen adsorption, pronounced spin redistribution occurs near the Si$_4$--N$_{16}$ region, indicating strong electronic coupling between the adsorbed hydrogen atom and the Si--N framework. 
			The spin depletion around Si$_4$ and accumulation around H highlight charge transfer processes associated with the Volmer step of the hydrogen evolution reaction (HER).}
		\label{spin_density}
	\end{figure}
	
	The spin density distribution, obtained from spin-polarized DFT calculations, was analyzed to identify the spatial localization of unpaired electrons. The spin density represents the difference between spin-up ($\alpha$) and spin-down ($\beta$) electron densities, i.e., $\rho_{\text{spin}} = \rho(\alpha) - \rho(\beta)$. A positive spin density indicates regions dominated by spin-up electrons, while negative values correspond to spin-down electron accumulation. This distribution thus provides a quantitative measure of the presence and localization of unpaired electrons within the system.
	The isosurface plot of the spin density is shown in Fig.~\ref{spin_density}, where yellow and cyan lobes represent regions of spin-up and spin-down electron accumulation, respectively. Significant spin localization is observed [Fig.~\ref{spin_density}(a)] on Si atoms (Si$_1$, Si$_3$, and Si$_4$) and adjacent nitrogen atoms (N$_{13}$, N$_{15}$, and N$_{16}$), indicating the presence of unpaired electrons essential for proton-coupled electron transfer (PCET), a key step in the hydrogen evolution reaction (HER). The pronounced spin polarization around these atoms suggests that they are electronically active and capable of participating in redox transformations.
	
	The Si--N moieties exhibiting high spin density are identified as the primary HER active sites. Si atoms, as centres of unpaired electron density, facilitate electron donation to incoming protons, forming surface-bound hydrogen intermediates (H*). Concurrently, adjacent N atoms stabilize adsorbed protons via electrostatic or hydrogen-bonding interactions. This cooperative mechanism supports the Volmer step (H$^+$ + e$^-$ $\rightarrow$ H*), followed by either the Tafel (2H* $\rightarrow$ H$_2$) or Heyrovsky (H* + H$^+$ + e$^-$ $\rightarrow$ H$_2$) pathway. The geometric and electronic configuration of the Si--N units thus provides a favourable environment for hydrogen adsorption and recombination. In contrast, negligible spin density on Sn atoms suggests their role is primarily structural or electronic modulation rather than direct HER participation.
	
	These comparison of spin density plots before and after H adsorption (Fig.~\ref{spin_density}) reveals notable redistribution of spin polarization. In the pristine structure [Fig.~\ref{spin_density}(a)], spin density is primarily localized around Si--N, designating these moieties as potential catalytic centres. Upon hydrogen adsorption [Fig.~\ref{spin_density}(b)], spin density concentrates near the N$_{16}$ region, with partial depletion on Si$_4$ and accumulation around the adsorbed H atom, confirming charge transfer from the Si--N framework to the hydrogen atom and indicating Volmer step initiation.
	
	The overlapping spin isosurfaces between Si$_4$ and N$_{16}$ imply strong electronic coupling and an active HER site capable of efficient proton--electron transfer. The Si atom(3p$^2$) donates electrons to the adsorbed proton, forming a Si--H bond, while the adjacent N atom stabilizes the intermediate through electrostatic or weak hydrogen-bonding interactions. This synergy lowers the kinetic barrier for hydrogen adsorption and subsequent H$_2$ evolution, whereas the minimal spin density on Sn atoms reaffirms their role as structural stabilizers rather than direct catalytic centres.
	
	Overall, the redistribution of spin density upon H adsorption confirms significant electron transfer between the Si and N atoms, indicating strong electronic coupling and stabilization of the adsorbed hydrogen. These observations provide clear electronic evidence of site IV-specific hydrogen interaction, supporting its catalytic performance. Collectively, these results highlight the 2D SnGeSiN$_4$ monolayer as a highly next-generation material for HER photocatalysis.
	
	\section{Conclusion}
	In this work, we have conducted a comprehensive first-principles investigation of the newly predicted two-dimensional SnSiGeN$_4$ MXene-family monolayer as a potential photocatalyst for water splitting. Structural optimization in the triclinic space group \textit{P1} confirms its dynamical and mechanical stability, while electronic band structures calculated using multiple exchange–correlation functionals reveal a different band gap. Hybrid functionals such as HSE06 and B3LYP provide more reliable estimates compared to semi-local approximations. 
	Vibrational and optical analyses show distinct Raman- and IR-active phonon modes along with pronounced in-plane and out-of-plane anisotropy, reflecting the layered nature of the system. The optical spectra exhibit strong absorption in the UV–visible region, supporting efficient photon harvesting and charge separation under solar illumination.
	Photocatalytic free-energy analyses for the OER, ORR, and HER indicate exceptionally low overpotentials. Notably, the OER overpotential at Site~V is 0.48~V, lower than that of the benchmark IrO$_2$ catalyst (0.56~V) and comparable to PtO$_2$-rutile (0.4~V). Gibbs free energy profiles and vibrational analysis identify Sites V and VI as the most active, while HER calculations reveal that Sites I, III, IV, and VI exhibit favourable hydrogen adsorption energies, with some sites surpassing the performance of Pt, MoS$_2$, and WS$_2$. These results highlight not only the high catalytic efficiency of SnSiGeN$_4$ but also its potential as a cost-effective alternative to scarce and expensive noble-metal catalysts.
	Overall, SnSiGeN$_4$ emerges as a stable, sustainable, and high-performance two-dimensional photocatalyst for next-generation UV–visible-light-driven water splitting. These theoretical insights provide a strong foundation for future experimental validation and practical application in clean energy technologies.
	
	\begin{acknowledgement}
		\textbf{DPR} acknowledges Anusandhan National Research Foundation (ANRF), a statutory body of the Department of Science \& Technology (DST), Government of India for the project Sanction Order No.: CRG/2023/000310, dated: 10 October, 2024.  Computational resources provided by the Paramrudra High-Performance Computing (HPC) facility at the Inter-University Accelerator Centre (IUAC), New Delhi.\\
		\textbf{CBS} acknowledge fruitful discussions with Shrish Nath Upadhyay from IIT Kanpur, India.\\
		\textbf{Amel Laref} acknowledges support from the "Research Center of the Female Scientific and Medical Colleges",  Deanship of Scientific Research, King Saud University.\\ 
	\end{acknowledgement}
	%The research is partially funded by the Ministry of Science and Higher Education of the Russian Federation as part of the World-Class Research Center program: Advanced Digital Technologies (contract No. 075-15-2022-312 dated 20.04.2022).
	
	\section*{Author contributions}
	\begin{itemize}
		\item \textbf{Chhatra Bahadur Subba:} First author, Formal analysis, Visualization, Validation, survey of literature, Writing-original draft, writing-review \& editing.
		\item \textbf{Bhanu Chettri}:Formal analysis, Visualization, Validation, writing-review \& editing. 
		\item \textbf{Amel Laref}:Formal analysis, Visualization, Validation, writing-review \& editing. 
		\item \textbf{Zeesam Abbas}:Formal analysis, Visualization, Validation, writing-review \& editing. 
		\item \textbf{Amna Parveen}:Formal analysis, Visualization, Validation, writing-review \& editing. 
		\item \textbf{Dibya Prakash Rai:} Project management, Supervision, Resources, Formal analysis, Visualization, Validation, writing-review \& editing. 
		\item \textbf{Zaithanzauva Pachuau}:Formal analysis, Visualization, Validation, writing-review \& editing. 
	\end{itemize}
	
	\section*{Data Availability Statement}
	Data is available in the article.

	\begin{suppinfo}
		
	\end{suppinfo}
	
	\bibliography{achemso-demo}
	
\end{document}